# Modeling Unsteady Aircraft Aerodynamics Using Lorenz Attractor: A Reduced-Order Approach for Wing Rock

Marcel Menner[1]
*Aurora Flight Sciences (A Boeing Company), Cambridge, MA, 02142, USA*

Eugene Lavretsky[2]
*The Boeing Company, Huntington Beach, CA, 92647, USA*

**This paper presents a novel modeling approach for unsteady aircraft airflow, leveraging the Lorenz attractor framework. The proposed model is based on the force distribution exerted by a lift-generating wing on the surrounding fluid. It distinguishes between turbulent and nominal components of the force distribution, with the nominal force distribution modeled to peak at the wing and decay linearly into the free stream. This separation allows the turbulent component to be represented by a transport equation that is influenced by flight conditions, specifically dynamic pressure and angle of attack. Consequently, the Navier-Stokes equations, along with the turbulence transport equation, can be transformed into a reduced-order model characterized by three scalar ordinary differential equations – similar to the Lorenz attractor. This resulting system effectively captures chaotic behavior, facilitating the exploration of complex dynamics without the computational demands of solving the full Navier-Stokes equations. A simulation trade study is conducted that models wing rock phenomena at high angles of attack, demonstrating the effectiveness of the proposed approach in capturing the intricate dynamics of unsteady aircraft aerodynamics.**

## I. Introduction

The study of unsteady flow around aircraft wings is a vital aspect of aerodynamics, with significant implications for aircraft design and performance. A comprehensive understanding of airflow dynamics, especially during high angles of attack, is crucial for accurately predicting lift, drag, and overall aerodynamic efficiency. Traditional modeling approaches typically involve solving the Navier-Stokes equations, which, while yielding precise results, can be computationally demanding. In this context, the Lorenz attractor framework offers a promising alternative by effectively capturing complex and chaotic dynamics within a reduced-order model. This research is driven by the need to develop a reduced-order model for wing rock, a dynamic instability phenomenon that occurs in aircraft, particularly at high angles of attack. Wing rock is characterized by oscillatory motion of the aircraft's wings, where one wing experiences a sudden increase in lift while the other wing experiences a decrease. This lift imbalance can induce a rolling motion, causing the aircraft to rock back and forth around its longitudinal axis [1]. Understanding the mechanisms behind wing rock is crucial for improving aircraft stability and performance.

We propose a novel modeling technique that decomposes the force distribution exerted by a lift-generating wing into nominal and turbulent components. The nominal component is modeled to peak at the wing and decay linearly into the free stream, while the turbulent component is represented by a transport equation influenced by flight conditions such as dynamic pressure and angle of attack. By employing a Galerkin-Fourier framework akin to the Lorenz attractor, this separation enables a more efficient representation of airflow characteristics. Utilizing this model, we can simplify the Navier-Stokes equations and the turbulence transport equation into a reduced-order system characterized by three ordinary differential equations. This abstraction allows for the exploration of complex dynamics without the computational burden associated with solving the full Navier-Stokes equations. Our simulation trade study demonstrates the model's effectiveness in predicting flow separation at high angles of attack, establishing it as a valuable tool for modeling wing rock and enhancing real-time simulations and flight control design. The findings underscore the potential of the Lorenz attractor-based approach to improve performance and safety of aircraft experiencing unsteady airflow [1].

---
[1] Senior Engineer, Autonomy.
[2] Senior Principal Technical Fellow, Boeing Research & Technology. AIAA Fellow.

**A. Related Works**

Reduced-order modeling has emerged as a powerful tool in the field of aerodynamics, particularly for capturing the complex dynamics of turbulent flows around airfoils. Traditional CFD methods, while accurate, often require significant computational resources, making them less suitable for real-time applications [2], [3], [4]. To address this challenge, various reduced-order modeling techniques that simplify the governing equations while retaining essential flow characteristics have been deployed. For instance, the Proper Orthogonal Decomposition (POD) method has been widely employed to extract dominant modes from high-fidelity simulations, allowing for a more efficient representation of turbulent flows [5]. Dynamic Mode Decomposition (DMD) provides a framework for analyzing the temporal evolution of unsteady flow structures, enabling the identification of dominant modes and their associated dynamics [6]. Recent studies have shown that DMD can effectively capture the dynamics of flow separation and reattachment [7]. A popular alternative is the use of machine learning techniques to enhance traditional reduced-order modeling approaches. E.g., data-driven approaches such as neural networks have been employed to approximate the mapping between flow states and aerodynamic forces [8]. Turbulence modeling remains a critical aspect of aerodynamic simulations to predict flow behavior around airfoils, where Large Eddy Simulation (LES) and Reynolds-Averaged Navier-Stokes (RANS) methods are commonly used [9]. Similar to our approach, RANS models rely on empirical closure models that aim to capture turbulent flows [10]. Here, too, the choice of turbulence model depends on the specific application, flow characteristics, and required accuracy [4].

The research studying wing rock often focuses on both control design and modeling using computational fluid dynamics (CFD). E.g., in [11], an L1 adaptive controller is designed to suppress wing-body rock, and [12] performed a nonlinear analysis of the wing-rock system with adaptive control. The research in [11], [12] highlights the potential for advanced control strategies to mitigate wing rock effects. The research in [13] focused on predicting wing-rock limit cycle oscillations using CFD, providing valuable insights into the aerodynamic factors contributing to this phenomenon. Numerical simulations of free-to-roll wing rock phenomena are presented in [14], utilizing time spectral CFD techniques to analyze the dynamics. Additionally, [15] investigated wing rock in a fighter aircraft through free-to-roll wind tunnel tests and dynamic CFD methods. Further, control strategies for unsteady flows are also researched in the literature on underwater vehicles, which can experience unsteady hydrodynamic loads [16], [17]. Different from [11]– [17], we propose a reduced-order model to predict wing rock and show how control systems can adapt to these instabilities in real-time by incorporating a model based on the Lorenz dynamics into the controller.

## II. Mathematical Background

We build upon the foundational work presented by Salzman [18], which models fluid motion between two fixed boundaries using a convection-diffusion transport equation driven by a nominal temperature gradient. In [18], a Galerkin-Fourier expansion is employed with a Boussinesq–Oberbeck approximation [19], demonstrating that the resulting fluid dynamics can be effectively captured by a limited number of dominant basis functions. Lorenz [20] further showed that the same result can be obtained by reducing the Navier-Stokes and transport equation to those dominating basis function in combination with the Lorenz attractor, which models deterministic nonperiodic flow.

To this end, consider the set of differential equations of the form [18]

$$\frac{\partial}{\partial t} \nabla^2 \Psi(x,z,t) = -\frac{\partial(\Psi(x,z,t), \nabla^2 \Psi(x,z,t))}{\partial(x,z)} + \nu \nabla^4 \Psi(x,z,t) + g\alpha \frac{\partial \xi(x,z,t)}{\partial x}$$
$$\frac{\partial}{\partial t} \xi(x,z,t) = -\frac{\partial(\Psi(x,z,t), \xi(x,z,t))}{\partial(x,z)} + \frac{\Delta T}{H} \frac{\partial \Psi(x,z,t)}{\partial x} + \kappa \nabla^2 \xi(x,z,t)$$

(1)

with the gravity $g$, the coefficient of thermal expansion $\alpha$, the kinematic viscosity $\nu$, and the thermal conductivity $\kappa$. In (1), $\xi(x,z,t)$ models the deviation of the temperature in the fluid from a nominal linear component, $\Psi(x,z,t)$ is a stream function with the boundary conditions

$$\left.\frac{\partial \Psi(x,z,t)}{\partial x}\right|_{z=0} = u_z(x, z=0, t) = 0, \qquad \left.\frac{\partial \Psi(x,z,t)}{\partial x}\right|_{z=H} = u_z(x, z=H, t) = 0, \tag{2}$$

$H$ is the characteristic length scale defining the distance between the two boundaries, $\nabla = \begin{bmatrix} \partial_x & \partial_y & \partial_z \end{bmatrix}^T$ is the vector differential operator with the notation $\partial_x = \frac{\partial}{\partial x}$, and

$$\frac{\partial(\eta(x,z,t), \lambda(x,z,t))}{\partial(x,z)} = \frac{\partial \eta(x,z,t)}{\partial x}\frac{\partial \lambda(x,z,t)}{\partial z} - \frac{\partial \lambda(x,z,t)}{\partial x}\frac{\partial \eta(x,z,t)}{\partial z}. \tag{3}$$

In (2), the boundary conditions imply that the fluid can only move parallel to the boundaries.

Lorenz [20] consequently used the following basis functions derived from (1),

$$\Psi(x,z,t) = X(t)\frac{\kappa(1+a^2)\sqrt{2}}{a}\sin\left(\frac{\pi a}{H}x\right)\sin\left(\frac{\pi}{H}z\right)$$

$$\xi(x,z,t) = Y(t)\frac{\Delta T R_c \sqrt{2}}{R_a \pi}\cos\left(\frac{\pi a}{H}x\right)\sin\left(\frac{\pi}{H}z\right) - Z(t)\frac{\Delta T R_c}{R_a \pi}\sin\left(\frac{2\pi}{H}z\right) \tag{4}$$

with $R_c = \pi^4(1+a^2)^3/a^2$, $R_a = g\alpha H^3 \Delta T/(\nu\kappa)$, and a constant $a$, where $X(t)$, $Y(t)$, and $Z(t)$ reflect the time-dependent amplitudes and are governed by the three differential equations – called the Lorenz attractor,

$$\begin{aligned}
\frac{\partial X}{\partial \tau} &= \sigma(Y(t) - X(t)) \\
\frac{\partial Y}{\partial \tau} &= -X(t)Z(t) + RX(t) - Y(t) \\
\frac{\partial Z}{\partial \tau} &= X(t)Y(t) - bZ(t).
\end{aligned} \tag{5}$$

with the dimensionless time $\tau = \pi^2(1+a^2)\kappa t/H^2$, the Prandtl number $\sigma = \nu/\kappa$, $R = R_a/R_c$, and $b = 1/(1+a^2)$.

### III. Aircraft Aerodynamics and Lift Modeling

In this paper, we propose a model for the force and flow field of an aircraft, which is built on the separation of nominal and turbulence-induced components. Using a set of simplifying assumptions, our goal is to derive a reduced-order model that can be used for real-time simulation and active control of an aircraft whose behavior changes with turbulent flow as a function of flight conditions.

Consider a nominal force distribution that the wing imposes on the surrounding fluid illustrated in Fig. 1 with a wing span $s$ and an airfoil with chord length $c$. Of primary interest is the vertical force component that the fluid imposes on the wing as it relates to lift. The force distribution is assumed to (i) have a peak $\Delta L$ at the edge of the wing, (ii) vanish at a distance $H$ (free stream), and (iii) be linear in between. In this model, the forces on the fluid are positive, i.e., the fluid is being accelerated along the positive $z$ axis, which causes lift for the aircraft based on Newton's third law of motion. Since the model assumes symmetry between the force distribution above and below the wing, we focus only on the fluid behavior above the wing. While being a simplification of the true force distribution of a wing, this model enables a reduced-order model with only three scalar differential equations similar to the Lorenz attractor, which is shown in this paper. Note that we use a different definition for the coordinate system compared to [20] in order to align the axes with the body-fixed system commonly used in flight controls.

Using the model for the nominal forces in Fig. 1, we define the location and time-dependent force distribution governing lift and airflow accelerations $F_z(y,z,t)$, valid for the fluid above the wing,

$$\begin{aligned}
F_z(y,z,t) &= F_{\text{nom}}(z) - \xi(y,z,t) \\
F_{\text{nom}}(z) &= \Delta L + \frac{\Delta L}{H}z
\end{aligned} \tag{6}$$

with $F_{\text{nom}}(z,\alpha,\bar{q})$ defining the nominal force distribution and $\xi(y,z,t)$ is used to account for turbulent effects, unsteady aerodynamics, or flow separation. Similar to [20], $\xi(y,z,t)$ thus defines the deviation from a nominal linear profile, and is governed by the Navier-Stokes equation with a turbulence transport equation, which is discussed in detail in Section IV. The reason for the proposed model in Fig. 1 is to build on the research in [20] and adapt its results to unsteady aircraft aerodynamics.

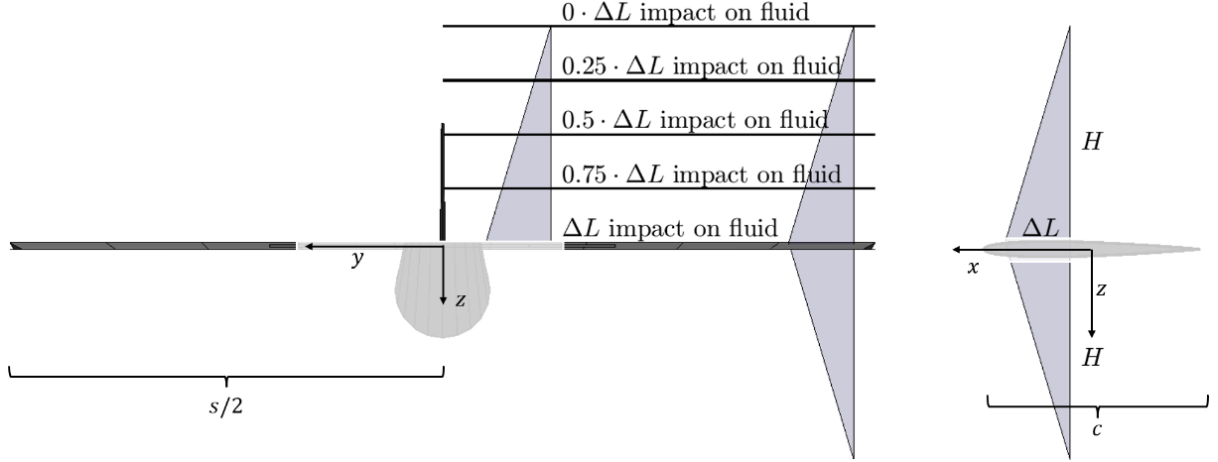

**Fig. 1. Model of Nominal Force Impact on Fluid due to Lift**

The nominal lift component in (6) of the aircraft in the absence of turbulence or flow separation can be calculated using the volume integral with

$$L_{\text{nom}} = \oiiint_{V_{\text{wing}}} F_{\text{nom}}(z) dV = \int_{y=-s/2}^{s/2} \int_{z=-H}^{0} \int_{x=-c/2}^{c/2} \left(\Delta L + \frac{\Delta L}{H} z\right) dxdzdy + \int_{y=-s/2}^{s/2} \int_{z=0}^{H} \int_{x=-c/2}^{c/2} \left(\Delta L - \frac{\Delta L}{H} z\right) dxdzdy$$
$$= sHc\frac{\Delta L}{2} + Hc\frac{\Delta L}{2} = sHc\Delta L, \quad (7)$$

where $dV = dxdydz$ and the closed volume as illustrated in Fig. 1 is given by

$$V_{\text{wing}} = \left\{(x,y,z): x \in \left[-\frac{c}{2},\frac{c}{2}\right], y \in \left[-\frac{s}{2},\frac{s}{2}\right], z \in [-H,H]\right\} \quad (8)$$

with the chord length $c$, the wing span $s$, and the characteristic length $H$ defining the distance from the wing to free stream conditions. Then, in accordance with Newton's third law of motion, the force distribution in (7) can be associated with flight conditions

$$L_{\text{nom}}(\alpha,\bar{q}) = \frac{1}{2} C_L(\alpha) \rho A V_\infty^2 = \bar{q} A C_L(\alpha) = sHc\Delta L = AH\Delta L \quad (9)$$

with the wing surface area $A = sc$, which implies

$$\Delta L = \Delta L(\alpha,\bar{q}) = \frac{C_L(\alpha)\rho V_\infty^2}{2H} = \frac{\bar{q} C_L(\alpha)}{H} \quad (10)$$

with the aircraft's airspeed $V$, the lift coefficient $C_L(\alpha)$ as a function of angle of attack (AOA) $\alpha$, and dynamic pressure $\bar{q} = \rho V_\infty^2 / 2$, where $\Delta L = \Delta L(\alpha,\bar{q})$ is used to emphasize the dependence of $\Delta L$ on the flight conditions.

**Remark 1.** While we use a force distribution, which is constant across the wing surface, this is not a necessary assumption in practice. Instead, $\Delta L$ can represent a mean value of a varying force distribution that is used to derive an expression for the turbulence-induced force component $\xi(y,z,t)$. The primary motivation for this research is the development of a reduced-order model for wing rock that can be used for efficient simulation and control design.

# IV. Modeling of Unsteady Flow using Lorenz Attractor

We model the fluid motion using three nonlinear partial differential equations (PDEs), with the continuity equation

$$\nabla \cdot u(y,z,t) = 0, \qquad (11)$$

the incompressible Navier-Stokes equation

$$\frac{\partial u(y,z,t)}{\partial t} + (u(y,z,t) \cdot \nabla)u(y,z,t) = -\frac{1}{\rho}\nabla p(y,z,t) + v_\omega \Delta u(y,z,t) + \frac{1}{\rho}F(y,z,t)$$

$$F(y,z,t) = \begin{bmatrix} 0 \\ 0 \\ F_z(y,z,t) \end{bmatrix}, \qquad (12)$$

and a scalar transport equation modeling the conservation of turbulent momentum

$$\frac{\partial \xi(y,z,t)}{\partial t} + (u(y,z,t) \cdot \nabla)\xi(y,z,t) + (u(y,z,t) \cdot \nabla)\frac{\Delta L(\alpha,\bar{q})}{H}z = v_\phi \Delta \xi(y,z,t). \qquad (13)$$

In (12) and (13), $p$ is the pressure field, $v_\omega$ is the kinematic viscosity, $\rho$ is the fluid density, $F$ is the force field, $v_\phi$ is the turbulent diffusion coefficient, $P = (u \cdot \nabla)\Delta Lz/H$ is a production term resulting from the nominal force field in (7), and $\Delta = \nabla \cdot \nabla = \begin{bmatrix} \partial_x^2 & \partial_y^2 & \partial_z^2 \end{bmatrix}^T$ is the Laplace operator.

*Remark 2.* In this model, the production term—and consequently the flight conditions—drives turbulence.

*Remark 3.* Eq. (13) represents a transport equation, which is related to turbulence models in CFD such as $k-\epsilon$ [21], $k-\omega$ [22], or Reynolds stress equation model [23]. E.g., in a $k-$model, $\xi$ is related to turbulent kinematic energy. While in such a CFD model, the kinematic turbulent eddy viscosity $v_\phi$ itself is often governed by differential equations, we show that an explicit solution with only three scalar ordinary differential equations can be derived using our modeling assumption with a constant/averaged $v_\phi$. Note that the kinematic turbulent eddy viscosity $v_\phi$ is considerable larger than the kinematic viscosity of a fluid.

## A. Reduced-Order Model

As we are primarily interested in asymmetric lift and its resulting roll moment, we model the airflow variations in the $y-z$ plane only with fluid velocity

$$u(y,z,t) = \begin{bmatrix} V_\infty \\ u_y(y,z,t) \\ u_z(y,z,t) \end{bmatrix}, \qquad (14)$$

with constant airspeed $V_\infty$. As we will show in this paper, these simplifications can be used to derive a Lorenz attractor model similar to (4) with (5) from the governing PDEs (12) and (13) to model flight condition-dependent airflow. The boundary conditions in this model imply

$$u_z(y,z,t)\big|_{z=0} = 0$$
$$u_z(y,z,t)\big|_{z=-H} = 0, \qquad (15)$$

where the first boundary condition is due to the wing and the second boundary condition is due to the free stream assumption.

Similar to [20], we define a stream function $\Psi(y,z,t)$ and the deviation from the nominal force distribution $\xi(y,z,t)$ in (6) using a Galerkin-Fourier expansion. In our model, fluid motion is induced by a lift/force gradient, whereas [20] studies fluid motion induced by a temperature gradient. However, since we also use a linear nominal distribution, we can build on the findings in [20] and define

$$\Psi(y,z,t) = X(t)\mathcal{A}_x \sin\left(\frac{\pi a}{H}(y-y_0)\right)\sin\left(-\frac{\pi}{H}z\right)$$

$$\xi(y,z,t) = Y(t)\mathcal{A}_y \cos\left(\frac{\pi a}{H}(y-y_0)\right)\sin\left(-\frac{\pi}{H}z\right) - Z(t)\mathcal{A}_z \sin\left(-\frac{2\pi}{H}z\right), \quad (16)$$

where the parameters $(\mathcal{A}_x, \mathcal{A}_y, \mathcal{A}_z)$ depend on the flight conditions and $(a, y_0)$ relate to characteristic length scales and are used to adapt the stream function and deviation from the nominal force distribution to the wing characteristics. The fluid velocities resulting from the stream function are given by

$$u_y(y,z,t) = -\frac{\partial \Psi(y,z,t)}{\partial z} = X(t)\mathcal{A}_x \frac{\pi}{H}\sin\left(\frac{\pi a}{H}(y-y_0)\right)\cos\left(-\frac{\pi}{H}z\right)$$

$$u_z(y,z,t) = \frac{\partial \Psi(y,z,t)}{\partial y} = X(t)\mathcal{A}_x \frac{\pi a}{H}\cos\left(\frac{\pi a}{H}(y-y_0)\right)\sin\left(-\frac{\pi}{H}z\right) \quad (17)$$

and the time and location-dependent force distribution

$$F_z(y,z,t) = \Delta L(\alpha,\bar{q}) + \frac{\Delta L(\alpha,\bar{q})}{H}z - Y(t)\mathcal{A}_y \cos\left(\frac{\pi a}{H}(y-y_0)\right)\sin\left(-\frac{\pi}{H}z\right) + Z(t)\mathcal{A}_z \sin\left(-\frac{2\pi}{H}z\right), \quad (18)$$

Clearly, the fluid velocities in (17) satisfy the boundary conditions in (15), the continuity equation in (11), and are in the same form as (2). The time-dependent amplitudes are given by the three-state Lorenz dynamics, with parameters that depend on the flight condition

$$\dot{X}(t) = \frac{\partial X}{\partial \tau}\frac{\partial \tau}{\partial t} = \mathcal{A}_\tau \sigma (Y(t) - X(t))$$

$$\dot{Y}(t) = \frac{\partial Y}{\partial \tau}\frac{\partial \tau}{\partial t} = \mathcal{A}_\tau (-X(t)Z(t) + RX(t) - Y(t)) \quad (19)$$

$$\dot{Z}(t) = \frac{\partial Z}{\partial \tau}\frac{\partial \tau}{\partial t} = \mathcal{A}_\tau (X(t)Y(t) - bZ(t)).$$

Note that in [18], [20], a normalized time $\tau$ is used, which is captured by the constant term $\partial_t \tau = \mathcal{A}_\tau$. What remains is to derive the parameters $(\mathcal{A}_x, \mathcal{A}_y, \mathcal{A}_z)$ in (17) and (18) as well as $(\mathcal{A}_\tau, \sigma, R, b)$ in (19) based on the Navier-Stokes equation (12) and the turbulent momentum transport equation (13) tailored to the lift model proposed in Section III.

**B. Parameters derived from Navier-Stokes and Turbulence Transport Equations**

This section derives the parameters in (4) and (5) that provide a solution to our aircraft aerodynamics model with the Navier-Stokes equation in (12), based on the proposed force distribution model in (6). We use a formulation based on the curl $\omega(y,z,t)$ of the flow field,

$$\omega(y,z,t) = \nabla \times u(y,z,t) = \begin{bmatrix}\partial_x \\ \partial_y \\ \partial_z\end{bmatrix} \times \begin{bmatrix}V_\infty \\ u_y(y,z,t) \\ u_z(y,z,t)\end{bmatrix} = \begin{bmatrix}\partial_y u_z(y,z,t) - \partial_z u_y(y,z,t) \\ 0 \\ 0\end{bmatrix}, \quad (20)$$

since the fluid motion occurs in the $y-z$ plane only. We apply $\nabla \times$ to both sides of the Navier-Stokes equation, which leads to the following governing equation for the curl of the fluid field

$$\frac{\partial \omega(y,z,t)}{\partial t} + (u(y,z,t)\cdot\nabla)\omega(y,z,t) = v_\omega \Delta \omega(y,z,t) + \frac{1}{\rho}\left[\nabla \times F(y,z,t)\right] \quad (21)$$

with $F(y,z,t)$ in (6). Note that $\nabla \times (\nabla p) = 0$. Appendix A shows that (21) is satisfied at all times and locations with the fluid velocities in (17) and the Lorenz attractor dynamics in (5) if the following conditions on the parameters hold

$$v_\omega \left(\frac{\pi}{H}\right)^2 (a^2+1) = \mathcal{A}_\tau \sigma, \quad \mathcal{A}_y = -\mathcal{A}_\tau \mathcal{A}_x \sigma \rho \frac{\pi}{H}\frac{(a^2+1)}{a}. \quad (22)$$

Next, consider the transport equation in (13) with $(u(y,z,t)\cdot\nabla)z = u_z(y,z,t)$,

$$\frac{\partial \xi(y,z,t)}{\partial t} + (u(y,z,t) \cdot \nabla)\xi(y,z,t) + \frac{\Delta L}{H}u_z(y,z,t) = v_\phi \Delta \xi(y,z,t). \tag{23}$$

Omitting trigonometric terms that do not explicitly appear in (16) – similar to the derivations in [20], the model parameters derived from (23) are given by

$$\frac{\partial \tau}{\partial t} = \mathcal{A}_\tau = v_\phi \left(\frac{\pi}{H}\right)^2 (a^2+1), \quad b = \frac{4}{(a^2+1)}, \quad R = -\frac{\Delta L a}{v_\omega \pi (a^2+1)}\frac{\mathcal{A}_x}{\mathcal{A}_y},$$

$$-\mathcal{A}_x \mathcal{A}_y \left(\frac{\pi}{H}\right)^2 a = 2\mathcal{A}_\tau \mathcal{A}_z, \quad -2\mathcal{A}_x \mathcal{A}_z \left(\frac{\pi}{H}\right)^2 a = \mathcal{A}_\tau \mathcal{A}_y, \tag{24}$$

which is shown in Appendix B.

Then, the conditions in (22) and (24) fully and uniquely specify the parameters of the proposed reduced-order model. Overall, the flight condition-dependent parameters are given by

$$b = \frac{4}{(a^2+1)}, \quad \sigma = \frac{v_\omega}{v_\phi}, \quad R = \Delta L \frac{a^2 H^3}{\rho v_\omega v_\phi \pi^4 (a^2+1)^3}, \quad \Delta L = \frac{\bar{q} C_L(\alpha)}{H}$$

$$\mathcal{A}_\tau = v_\phi \left(\frac{\pi}{H}\right)^2 (a^2+1), \quad \mathcal{A}_x = -v_\phi \frac{(a^2+1)}{a}, \quad \mathcal{A}_y = \rho v_\omega v_\phi \left(\frac{\pi}{H}\right)^3 \frac{(a^2+1)^3}{a^2}, \quad \mathcal{A}_z = \frac{1}{2}\rho v_\omega v_\phi \left(\frac{\pi}{H}\right)^3 \frac{(a^2+1)^3}{a^2}. \tag{25}$$

## V. Simulation Trade Study: Free-To-Roll Aircraft Dynamics

### A. Problem Setup

To illustrate the effects of unsteady aerodynamics on an aircraft and model wing rock, we study the aircraft roll dynamics

$$\dot{p}(t) = \frac{1}{I_x}\left(-c_p p(t) + \bar{q} A C_l l \delta_{ail}(t) + M_{roll}(t)\right) \tag{26}$$

with roll rate $p(t)$, aileron control command $\delta_{ail}(t)$, inertia $I_x = 3000 \text{kg} \cdot \text{m}^2$, damping $c_p = 100 \text{N} \cdot \text{m} \cdot \text{s}/\text{rad}$, $C_l = 0.1$, $\rho = 1.225 \text{kg}/\text{m}^3$, $s = 10\text{m}$, $c = 3\text{m}$, $A = 30\text{m}^2$, and $l = 5\text{m}$ being the distance from the center of gravity to the aileron. We study an aircraft with airspeed as $V = 70\text{m}/\text{s}$ ($\bar{q} = 3001.25 \text{N}/\text{m}^2$) and the lift coefficient given by $C_L(\alpha) = 0.5 + C_{L\alpha}\alpha$ with $C_{L\alpha} = 5$ and the aircraft's angle of attack (AOA) $\alpha$. The characteristic length scale values are chosen as $y_0 = s/4$ and $a = 2H/s$, which make the trigonometric functions align with the edges of the wing with $u_z(y=-s/2,z,t) = u_z(y=s/2,z,t) = 0$. We initialize the Lorenz dynamics in (19) with $X(t=0) \sim \mathcal{N}(0,0.01)$, $Y(t=0) \sim \mathcal{N}(0,0.01)$, and $Z(t=0) \sim \mathcal{N}(0,0.01)$. For the free-to-roll aircraft dynamics, we compute the roll moment using the incremental change in angle of attack caused by the fluid's rotation, with the incremental lift force

$$\delta F_z(y,z,t) = \bar{q} C_{L\alpha} \delta \alpha(y,z,t) = \bar{q} C_{L\alpha} \frac{u_z(y,z,t)}{V_\infty}. \tag{27}$$

Hence, the roll moment is given by

$$M_{roll}(t) = \frac{1}{-H}\int_{z=0}^{-H}\int_{y=-s/2}^{s/2} y \cdot \delta F_z(y,z,t) \, dy \, dz = \frac{2s\bar{q}C_{L\alpha}}{\pi V_\infty}\mathcal{A}_x X(t) = c_{pX}X(t) \tag{28}$$

with $\mathcal{A}_x$ in (25). Therefore, the augmented system dynamics with the Lorenz states is given by

$$\dot{p}(t) = \frac{1}{I_x}\left(-c_p p(t) + \bar{q}AC_l l\delta_{\text{ail}}(t) + \underbrace{c_{pX} X(t)}_{\text{fluid vorticity inducing a/c roll moment}}\right)$$
$$\dot{X}(t) = \mathcal{A}_\tau \sigma(Y(t) - X(t))$$
$$\dot{Y}(t) = \mathcal{A}_\tau (-X(t)Z(t) + RX(t) - Y(t)) \quad (29)$$
$$\dot{Z}(t) = \mathcal{A}_\tau (X(t)Y(t) - bZ(t)),$$

and

$$c_{pX} = \frac{2s\bar{q}C_{L\alpha}}{\pi V_\infty} \mathcal{A}_x. \quad (30)$$

We use a proportional-integral (PI) controller designed using a Linear Quadratic Regulator (LQR) to stabilize the bank angle $\phi(t)$ with $\dot{\phi}(t) = p(t)$ and the integrated error $e_{I,\phi}(t)$ [24],

$$\delta_{\text{ail}}(t) = -1 e_{I,\phi}(t) - 1.771\phi(t) - 1.068 p(t)$$
$$\dot{e}_{I,\phi}(t) = \phi(t). \quad (31)$$

The parameters in (18) and (19) are chosen as $H = 8\text{m}$, $\sigma = 3$, and $v_\phi = 2$. Here, $H = 8\text{m}$ approximates the distance from the wing to free stream, $\sigma = 3$ is calibrated to achieve the transition to turbulent flow at a given and known angle of attack, and $v_\phi = 2$ scales $\mathcal{A}_\tau = \partial\tau/\partial t$ and is calibrated to match the time constant of unsteady flow phenomena encountered in wing rock.

*Remark* **4**. For more accurate representation of specific applications, these parameters can be calibrated, e.g., with CFD or flight test data. The purpose of our reduced-order model is not to replace a CFD analysis but primarily to anticipate the occurrence of turbulent phenomena in flight and facilitate efficient simulations and analysis.

### B. Unsteady Aerodynamics as Function of Flight Conditions

First, we examine the flight conditions that induce turbulent and chaotic fluid motion. Fig. 2 illustrates the external physical quantities that drive the Lorenz dynamics model, specifically the AOA and dynamic pressure, both of which significantly influence the force profile $\Delta L(\alpha, \bar{q})$ and $R$ in (19). Fig. 2 shows the parameter $R$ in dashed black. The colormap highlights the variance of the Lorenz states as an indicator of turbulence and unsteady flow. This variance is calculated with constant AOA and dynamic pressure from the time history signals of the Lorenz states after transient effects have dissipated and a steady state or limit cycle has been established. Notably, it is observed that below a critical value of approximately $R = 18$, the Lorenz states remain in a steady state, exhibiting no variance. However, once this critical threshold is surpassed—either through an increase in the AOA or dynamic pressure—the Lorenz states transition into an unsteady pattern, indicating the onset of turbulence.

### C. Wing Rock Simulation Results

Next, we investigate the roll dynamics described in equation (26) across various angles of attack to illustrate the transition from laminar to turbulent flow modeled by means of the Lorenz attractor. The upper plot in Fig. 3 shows the variations of the AOA in this study. Specifically, at 5 seconds, an angle of attack of 5 degrees is commanded; this is followed by an increase to 15 degrees around 65 seconds, and a further increase to 25 degrees at 125 seconds. For AOA of 5 degrees and 15 degrees, the states of the Lorenz attractor converge rapidly to a steady-state value after the transient, characterized by a small magnitude, dissipates. This is expected from the result in Fig. 2 with the flight conditions $\bar{q} = 3001.25\text{N}/\text{m}^2$ and $\alpha = 15\text{deg}$. Consequently, the aircraft does not exhibit a significant bank angle. However, when the commanded angle of attack reaches 25 degrees, the airflow transitions to a turbulent and chaotic state. This chaotic behavior induces a roll moment that causes the aircraft to develop a significant bank angles of up to 20 degrees, despite being controlled by a PI controller as described in (31).

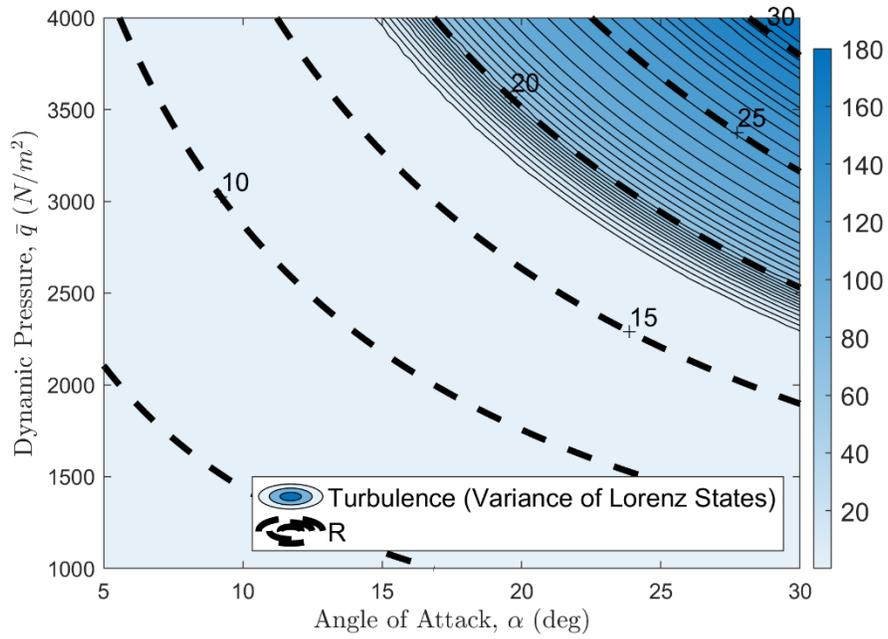

**Fig. 2. Flight Condition-dependent Turbulence**

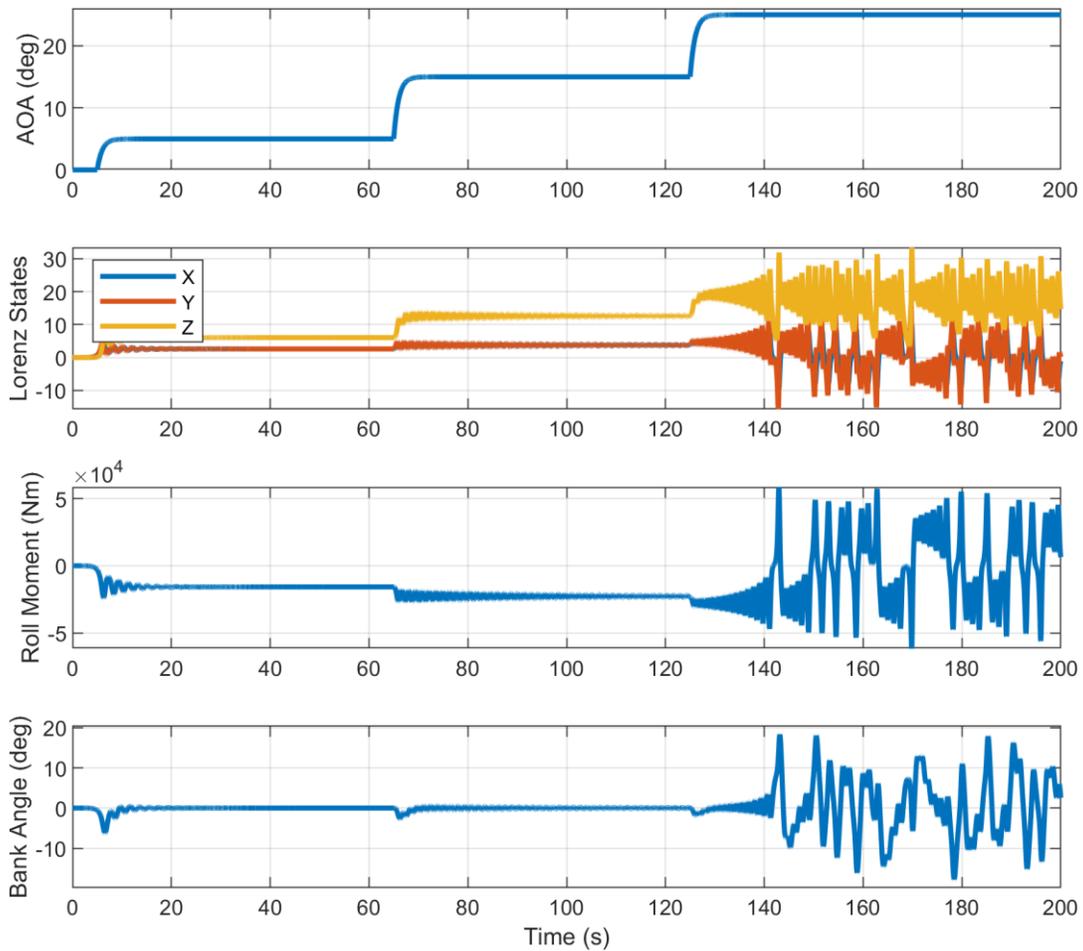

**Fig. 3. Onset of Unsteady Aerodynamic-induced Roll Moment at High Angle of Attack**

Fig. 4 provides a detailed examination of the high angle of attack scenario beginning at 160 seconds. The states of the Lorenz attractor appear chaotic, leading to alterations in the lift profiles. The resulting roll moment can produce bank angles of around 30 degrees. In this context, the Lorenz attractor model proves particularly valuable, as even minor variations in initial conditions can lead to substantial differences in the lift, which represents a useful framework for modeling turbulence.

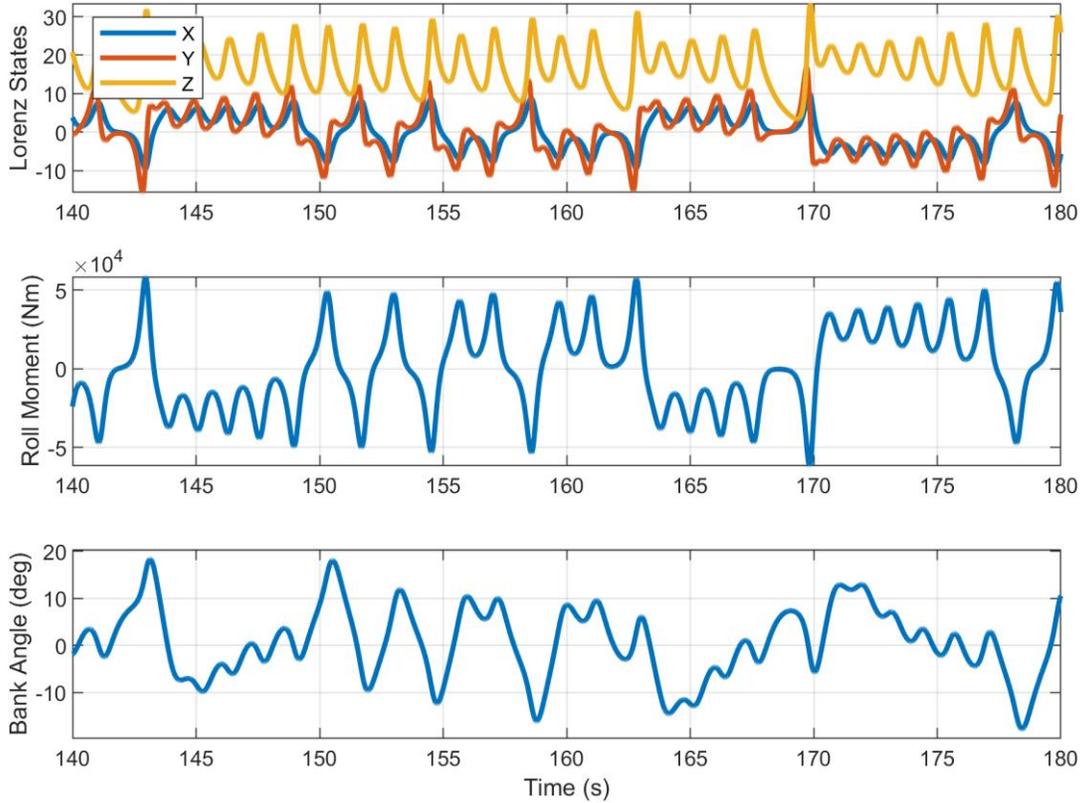

Fig. 4. Unsteady Aerodynamic-induced Roll Moment at 25deg Angle of Attack

### D. Force Distribution Simulation Results

Fig. 5 illustrates the total vertical force distribution $F_z(y,z,t)$, as described in equation (18), as a function of both location and time for the three different AOAs presented in Fig. 3. The colormap represents the magnitude of the force distribution. According to the proposed model, the wing is positioned at $z=0$. At an angle of attack of 5 degrees, the force distribution remains relatively constant over time, as evidenced in the top row of Fig. 5. At an angle of attack of 15 degrees, the force levels are generally elevated due to the increased lift associated with higher angles of attack; however, similar to the 5-degree case, the force distribution exhibits minimal temporal variation. At an angle of attack of 25 degrees, the force field demonstrates significant temporal fluctuations, as shown in the bottom row of Fig. 5. This variation in force distribution directly influences the total lift generated by the wing, contributing to the oscillations observed in Fig. 4.

Fig. 6 provides a more detailed examination of the force distribution, focusing on deviations from the nominal force distribution $\xi(y,z,t)$. Blue hues indicate a decrease in local force, whereas green hues indicate an increase. For the 5-degree angle of attack, the force distribution appears stable over time, with localized force values ranging between $-50\text{N}/\text{m}^3$ and $50\text{N}/\text{m}^3$. At 15 degrees of attack, the force field exhibits slight temporal changes, with values fluctuating between $-150\text{N}/\text{m}^3$ and $150\text{N}/\text{m}^3$. In the case of a 25-degree angle of attack, the force field experiences substantial temporal variation, with values spanning from $-500\text{N}/\text{m}^3$ to $500\text{N}/\text{m}^3$.

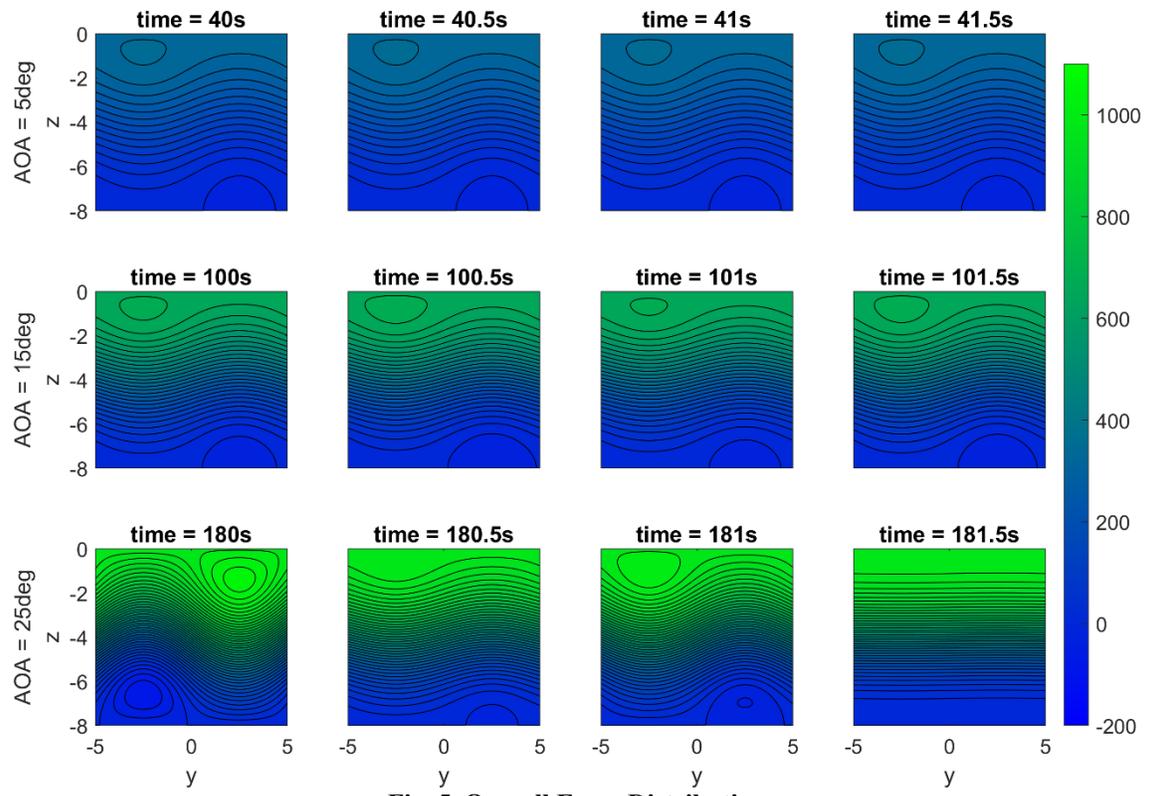

**Fig. 5. Overall Force Distribution**

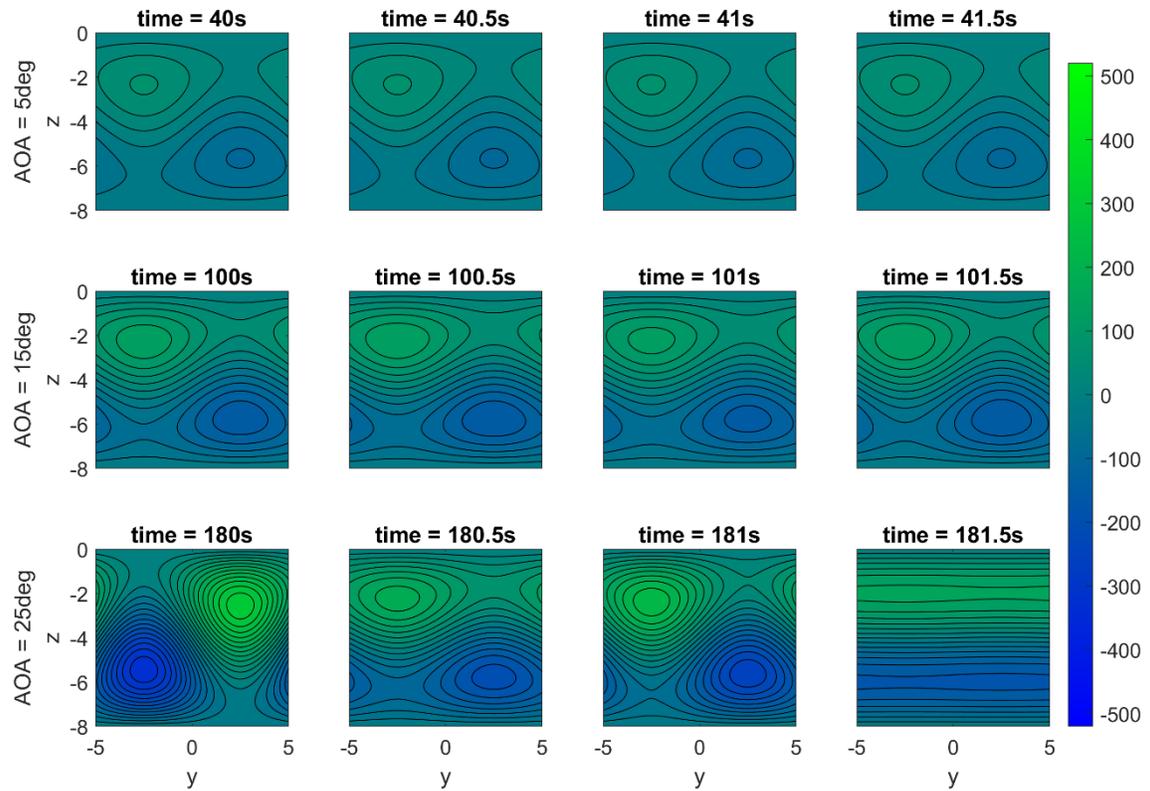

**Fig. 6. Deviation from Nominal Linear Force Profile**

### E. Turbulence-Augmented Controller Design

While the focus of this paper is on the modeling of turbulence using a reduced-order model, here we show the potential use for incorporating this model into the control design. Compared to turbulence models used in CFD, the proposed turbulence model can be executed in real time with negligible computational complexity.

Consider a modified PI controller, augmented with a turbulence feedback term,

$$\delta_{ail}(t) = \underbrace{-1 e_{I,\phi}(t) - 1.771\phi(t) - 1.068 p(t)}_{\text{Baseline}} + \underbrace{0.1 \hat{X}(t)}_{\substack{\text{Turbulence}\\\text{Augmentation}}}. \tag{32}$$

Since the states of the turbulence model in (19) are not directly measurable, $\hat{X}(t)$ in (32) is estimated using an Extended Kalman Filter with the following augmented dynamics model,

$$\begin{bmatrix} \dot{p}(t) \\ \dot{X}(t) \\ \dot{Y}(t) \\ \dot{Z}(t) \end{bmatrix} = \mathcal{A}_\tau \begin{bmatrix} -\dfrac{c_p}{I_x \mathcal{A}_\tau} & \dfrac{c_{pX}}{I_x \mathcal{A}_\tau} & 0 & 0 \\ 0 & -\sigma & \sigma & 0 \\ 0 & R - Z_t & -1 & -X_t \\ 0 & Y_t & X_t & -b \end{bmatrix} \begin{bmatrix} p(t) \\ X(t) \\ Y(t) \\ Z(t) \end{bmatrix} + \dfrac{1}{I_x} \begin{bmatrix} \bar{q} A C_l l \\ 0 \\ 0 \\ 0 \end{bmatrix} \delta_{ail}(t) \tag{33}$$

where $\{X, Y, Z\}_t$ result from linearization of the nonlinear system in (19) at time $t$ and the roll rate $p(t)$ is the only available measurement. Fig. 7 shows the difference between the baseline controller in (31) and the turbulence-augmented control design in (32). It shows that the turbulence augmentation in general reduced the bank angle and achieves bank angles of less than 10 degrees. While the control design cannot be expected to fully attenuate the seemingly chaotic aerodynamic effects, the turbulence augmentation reduced the deviation from zero degrees bank angle by around 72% on average, for the time duration illustrated in Fig. 7.

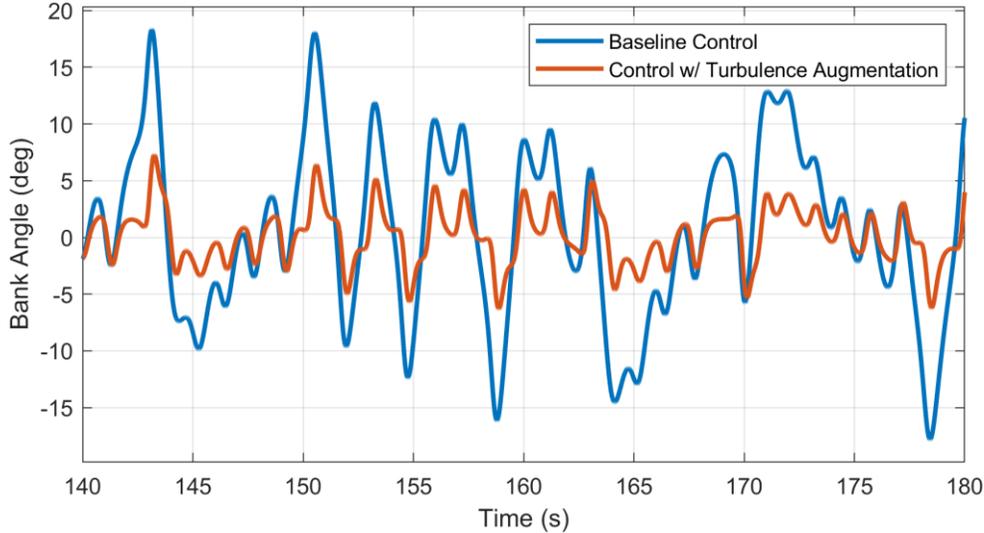

**Fig. 7. Reduction of Bank Angle by Turbulence-Augmented Control Design**

*Remark* **5**. It is easy to verify that the linearized system dynamics in (33) is observable for almost all states of the Lorenz dynamics [25]. Further, while the linearized system dynamics in (33) is not controllable, the nonlinear dynamics are stabilizable. These properties are the foundation of an advanced control design that uses the Lorenz dynamics.

*Remark* **6**. Note that (32) represents a simple solution for illustration purposes only. Future work will include a more sophisticated control design, which will improve performance further. For example, a more sophisticated control design may include the identification of dominant frequencies in the Lorenz dynamics and an appropriate attenuation instead of estimating the Lorenz states [24].

## VI. Conclusions

In this paper, we have presented a novel modeling approach for unsteady aerodynamic flows for aircraft, leveraging the Lorenz attractor framework to capture the complex dynamics of turbulence. By recognizing the relationship between the force distribution exerted by the wing and the resulting airflow, we developed a reduced-order model that simplifies the governing equations while retaining essential flow characteristics. This approach facilitates a more efficient representation of turbulent flows. The results of our simulation trade study demonstrate the effectiveness of the proposed model in capturing flow separation at high angles of attack, a critical aspect of aerodynamic performance. The ability to model chaotic behavior through a deterministic set of ordinary differential equations allows for the exploration of complex dynamics without the computational burden associated with full Navier-Stokes simulations. A simulation trade study showed the potential benefits of the proposed reduced-order model to predict wing rock and the potential for incorporating the proposed model into the flight control design. Future work will include a sophisticated control design that uses the proposed three-state turbulence model. Future work can also include a comparison with a higher-order Lorenz system [26].

## Appendix A: Derivation of Parameters in Navier-Stokes Equation

For brevity, we omit the dependence on time and location and use $y_0 = 0$ (without loss of generality) in the following. First, the velocities in (17) are used to derive the curl in (20),

$$\omega = \begin{bmatrix} \partial_y u_z - \partial_z u_y \\ 0 \\ 0 \end{bmatrix} = -X\mathcal{A}_x \left(\frac{\pi}{H}\right)^2 (a^2+1) \sin\left(\frac{\pi a}{H} y\right) \sin\left(-\frac{\pi}{H} z\right) \begin{bmatrix} 1 \\ 0 \\ 0 \end{bmatrix}. \tag{34}$$

Next, the individual terms in (21) are derived. Using (16), the curl of the force field in (6) is given by

$$\nabla \times F = \frac{\partial \left(\Delta L + \frac{\Delta L}{H} z - \xi\right)}{\partial y} \begin{bmatrix} 1 \\ 0 \\ 0 \end{bmatrix} = Y\mathcal{A}_y \frac{\pi a}{H} \sin\left(\frac{\pi a}{H} y\right) \sin\left(-\frac{\pi}{H} z\right) \begin{bmatrix} 1 \\ 0 \\ 0 \end{bmatrix}. \tag{35}$$

To obtain an expression for the material derivative,

$$(u \cdot \nabla)\omega = (u_y \partial_y + u_z \partial_z)\omega = X\mathcal{A}_x \frac{\pi}{H} \left( \sin\left(\frac{\pi a}{H} y\right) \cos\left(-\frac{\pi}{H} z\right) \partial_y + a \cos\left(\frac{\pi a}{H} y\right) \sin\left(-\frac{\pi}{H} z\right) \partial_z \right) \omega$$

$$= (X\mathcal{A}_x)^2 \left(\frac{\pi}{H}\right)^4 a(a^2+1) \sin\left(\frac{\pi a}{H} y\right) \cos\left(-\frac{\pi}{H} z\right) \cos\left(\frac{\pi a}{H} y\right) \sin\left(-\frac{\pi}{H} z\right) \begin{bmatrix} 1 \\ 0 \\ 0 \end{bmatrix} \tag{36}$$

$$- (X\mathcal{A}_x)^2 \left(\frac{\pi}{H}\right)^4 a(a^2+1) \cos\left(\frac{\pi a}{H} y\right) \sin\left(-\frac{\pi}{H} z\right) \sin\left(\frac{\pi a}{H} y\right) \cos\left(-\frac{\pi}{H} z\right) \begin{bmatrix} 1 \\ 0 \\ 0 \end{bmatrix} = 0.$$

The Laplace operator of the curl in (21) is given by

$$\Delta \omega = -(\partial_y^2 + \partial_z^2) X\mathcal{A}_x \left(\frac{\pi}{H}\right)^2 (a^2+1) \sin\left(\frac{\pi a}{H} y\right) \sin\left(-\frac{\pi}{H} z\right) \begin{bmatrix} 1 \\ 0 \\ 0 \end{bmatrix}$$

$$= X\mathcal{A}_x \left(\frac{\pi}{H}\right)^4 a^2(a^2+1) \sin\left(\frac{\pi a}{H} y\right) \sin\left(-\frac{\pi}{H} z\right) \begin{bmatrix} 1 \\ 0 \\ 0 \end{bmatrix} + X\mathcal{A}_x \left(\frac{\pi}{H}\right)^4 (a^2+1) \sin\left(\frac{\pi a}{H} y\right) \sin\left(-\frac{\pi}{H} z\right) \begin{bmatrix} 1 \\ 0 \\ 0 \end{bmatrix} \tag{37}$$

$$= X\mathcal{A}_x \left(\frac{\pi}{H}\right)^4 (a^2+1)^2 \sin\left(\frac{\pi a}{H} y\right) \sin\left(-\frac{\pi}{H} z\right) \begin{bmatrix} 1 \\ 0 \\ 0 \end{bmatrix}.$$

Finally, inserting (37), (36), (35) in the governing equation for the curl in (21), we get (for the $x-$ axis)

$$-\frac{\partial X}{\partial t}\mathcal{A}_x\left(\frac{\pi}{H}\right)^2(a^2+1)\sin\left(\frac{\pi a}{H}y\right)\sin\left(-\frac{\pi}{H}z\right)$$
$$=\nu_\omega X\mathcal{A}_x\left(\frac{\pi}{H}\right)^4(a^2+1)^2\sin\left(\frac{\pi a}{H}y\right)\sin\left(-\frac{\pi}{H}z\right)+\frac{1}{\rho}Y\mathcal{A}_y\frac{\pi a}{H}\sin\left(\frac{\pi a}{H}y\right)\sin\left(-\frac{\pi}{H}z\right),$$
(38)

which can equivalently be written as

$$\left(\frac{\partial X}{\partial t}\mathcal{A}_x\left(\frac{\pi}{H}\right)^2(a^2+1)+\nu_\omega X\mathcal{A}_x\left(\frac{\pi}{H}\right)^4(a^2+1)^2+\frac{1}{\rho}Y\mathcal{A}_y\frac{\pi a}{H}\right)\sin\left(\frac{\pi a}{H}y\right)\sin\left(-\frac{\pi}{H}z\right)=0.$$
(39)

Since (39) needs to be satisfied everywhere, the trigonometric expressions can be removed, and

$$\frac{\partial X}{\partial t}\mathcal{A}_x\left(\frac{\pi}{H}\right)^2(a^2+1)+\nu_\omega X\mathcal{A}_x\left(\frac{\pi}{H}\right)^4(a^2+1)^2+\frac{1}{\rho}Y\mathcal{A}_y\frac{\pi}{H}a=0.$$
(40)

We proceed by inserting the first differential equation in the Lorenz attractor

$$\frac{\partial X}{\partial \tau}\frac{\partial \tau}{\partial t}=\frac{\partial X}{\partial \tau}\mathcal{A}_\tau,\text{ with }\frac{\partial X}{\partial \tau}=\sigma(Y-X)$$
(41)

with $\frac{\partial \tau}{\partial t}=\mathcal{A}_\tau$ into (40),

$$X\mathcal{A}_x\frac{\pi}{H}(a^2+1)\left(\nu_\omega\left(\frac{\pi}{H}\right)^2(a^2+1)-\mathcal{A}_\tau\sigma\right)+Y\left(\mathcal{A}_\tau\sigma\mathcal{A}_x\frac{\pi}{H}(a^2+1)+\frac{1}{\rho}\mathcal{A}_y a\right)=0$$
(42)

Again, for (42) to be satisfied at all times, i.e., for all $X$ and $Y$, the necessary choice of parameters is given by (22).

## Appendix B: Derivation of Parameters in Turbulence Model

First, the time derivative in (23) with the Lorenz attractor dynamics is given by

$$\frac{\partial \xi}{\partial t}=\frac{\partial Y}{\partial \tau}\mathcal{A}_\tau\mathcal{A}_y\cos\left(\frac{\pi a}{H}y\right)\sin\left(-\frac{\pi}{H}z\right)-\frac{\partial Z}{\partial \tau}\mathcal{A}_\tau\mathcal{A}_z\sin\left(-\frac{2\pi}{H}z\right)$$
$$=(-XZ+RX-Y)\mathcal{A}_\tau\mathcal{A}_y\cos\left(\frac{\pi a}{H}y\right)\sin\left(-\frac{\pi}{H}z\right)-(XY-bZ)\mathcal{A}_\tau\mathcal{A}_z\sin\left(-\frac{2\pi}{H}z\right),$$
(43)

where we used the second and third differential equation in the Lorenz attractor (5),

$$\frac{\partial Y}{\partial \tau}\frac{\partial \tau}{\partial t}=\frac{\partial Y}{\partial \tau}\mathcal{A}_\tau,\text{ with }\frac{\partial Y}{\partial \tau}=-XZ+RX-Y$$
$$\frac{\partial Z}{\partial \tau}\frac{\partial \tau}{\partial t}=\frac{\partial Z}{\partial \tau}\mathcal{A}_\tau,\text{ with }\frac{\partial Z}{\partial \tau}=XY-bZ$$
(44)

Further, the second term in (23) yields

$$(u\cdot\nabla)\xi=X\mathcal{A}_x\frac{\pi}{H}\left(\sin\left(\frac{\pi a}{H}y\right)\cos\left(-\frac{\pi}{H}z\right)\partial_y+a\cos\left(\frac{\pi a}{H}x\right)\sin\left(-\frac{\pi}{H}z\right)\partial_z\right)\xi$$
$$=-X\mathcal{A}_x\frac{\pi}{H}\frac{\pi a}{H}\left(Y\mathcal{A}_y\cos\left(-\frac{\pi}{H}z\right)\sin\left(-\frac{\pi}{H}z\right)-2Z\mathcal{A}_z\cos\left(\frac{\pi a}{H}x\right)\sin\left(-\frac{\pi}{H}z\right)\cos\left(-\frac{2\pi}{H}z\right)\right)$$
$$=-X\mathcal{A}_x\frac{\pi}{H}\frac{\pi a}{H}\left(Y\mathcal{A}_y\cos\left(-\frac{\pi}{H}z\right)\sin\left(-\frac{\pi}{H}z\right)-2Z\mathcal{A}_z\cos\left(\frac{\pi a}{H}y\right)\sin\left(-\frac{\pi}{H}z\right)\left(2\cos^2\left(-\frac{\pi}{H}z\right)-1\right)\right),$$
(45)

where we used the two trigonometric identities

$$2\cos\left(\frac{\pi}{H}z\right)\sin\left(\frac{\pi}{H}z\right) = \sin\left(2\frac{\pi}{H}z\right), \quad \cos^2\left(\frac{\pi a}{H}z\right) + \sin^2\left(\frac{\pi a}{H}z\right) = 1. \tag{46}$$

As in [20], we omit trigonometric terms that do not explicitly appear in (16), and thus

$$(u \cdot \nabla)\xi = -X\mathcal{A}_x\left(\frac{\pi}{H}\right)^2 a\left(Y\mathcal{A}_y \frac{1}{2}\sin\left(-\frac{2\pi}{H}z\right) + 2Z\mathcal{A}_z \cos\left(\frac{\pi a}{H}y\right)\sin\left(-\frac{\pi}{H}z\right)\right), \tag{47}$$

Next,

$$\Delta\xi = \Delta\left(Y\mathcal{A}_y \cos\left(\frac{\pi a}{H}y\right)\sin\left(-\frac{\pi}{H}z\right) - Z\mathcal{A}_z \sin\left(-\frac{2\pi}{H}z\right)\right)$$
$$= -Y\mathcal{A}_y\left(\frac{\pi}{H}\right)^2 (a^2+1)\cos\left(\frac{\pi a}{H}y\right)\sin\left(-\frac{\pi}{H}z\right) + Z\mathcal{A}_z 4\left(\frac{\pi}{H}\right)^2 \sin\left(-\frac{2\pi}{H}z\right). \tag{48}$$

Finally, inserting the terms in (48), (47), and (43) into the transport equation (23) yields

$$(-XZ + RX - Y)\mathcal{A}_\tau\mathcal{A}_y \cos\left(\frac{\pi a}{H}y\right)\sin\left(-\frac{\pi}{H}z\right) - (XY - bZ)\mathcal{A}_\tau\mathcal{A}_z \sin\left(-\frac{2\pi}{H}z\right)$$
$$- X\mathcal{A}_x\left(\frac{\pi}{H}\right)^2 a\left(Y\mathcal{A}_y \frac{1}{2}\sin\left(-\frac{2\pi}{H}z\right) + 2Z\mathcal{A}_z \cos\left(\frac{\pi a}{H}y\right)\sin\left(-\frac{\pi}{H}z\right)\right)$$
$$+ \frac{\Delta L}{H} X\mathcal{A}_x \frac{\pi a}{H}\cos\left(\frac{\pi a}{H}y\right)\sin\left(-\frac{\pi}{H}z\right)$$
$$+ \nu_\phi Y\mathcal{A}_y\left(\frac{\pi}{H}\right)^2 (a^2+1)\cos\left(\frac{\pi a}{H}y\right)\sin\left(-\frac{\pi}{H}z\right) - \nu_\phi Z\mathcal{A}_z 4\left(\frac{\pi}{H}\right)^2 \sin\left(-\frac{2\pi}{H}z\right) = 0. \tag{49}$$

Finally, the necessary choice of parameters is obtained by satisfying (49) for all times and at all locations with

$$X\left(R\mathcal{A}_\tau\mathcal{A}_y + \frac{\Delta L}{H}\mathcal{A}_x \frac{\pi a}{H}\right)\cos\left(\frac{\pi a}{H}y\right)\sin\left(-\frac{\pi}{H}z\right) + Y\mathcal{A}_y\left(\nu_\phi\left(\frac{\pi}{H}\right)^2(a^2+1) - \mathcal{A}_\tau\right)\cos\left(\frac{\pi a}{H}y\right)\sin\left(-\frac{\pi}{H}z\right)$$
$$+ Z\left(b\mathcal{A}_\tau - \nu_\phi 4\left(\frac{\pi}{H}\right)^2\right)\mathcal{A}_z \sin\left(-\frac{2\pi}{H}z\right) \tag{50}$$
$$+ XY\left(-\mathcal{A}_x\left(\frac{\pi}{H}\right)^2 a\mathcal{A}_y \frac{1}{2} - \mathcal{A}_\tau\mathcal{A}_z\right)\sin\left(-\frac{2\pi}{H}z\right) + XZ\left(-\mathcal{A}_x\left(\frac{\pi}{H}\right)^2 a2\mathcal{A}_z - \mathcal{A}_\tau\mathcal{A}_y\right)\cos\left(\frac{\pi a}{H}y\right)\sin\left(-\frac{\pi}{H}z\right) = 0,$$

which leads to the choice of parameters in (24).